# Extracting and Analyzing Semantic Relatedness between Cities Using News Articles

Yingjie Hu [1], Xinyue Ye [2], and Shih-Lung Shaw [1]

[1] Department of Geography, University of Tennessee, Knoxville, TN 37996, USA
[2] Department of Geography, Kent State University, Kent, OH 44240, USA

**Abstract**. News articles capture a variety of topics about our society. They reflect not only the socioeconomic activities that happened in our physical world, but also some of the cultures, human interests, and public concerns that exist only in the perceptions of people. Cities are frequently mentioned in news articles, and two or more cities may co-occur in the same article. Such co-occurrence often suggests certain relatedness between the mentioned cities, and the relatedness may be under different topics depending on the contents of the news articles. We consider the relatedness under different topics as *semantic relatedness*. By reading news articles, one can grasp the general semantic relatedness between cities; yet, given hundreds of thousands of news articles, it is very difficult, if not impossible, for anyone to manually read them. This paper proposes a computational framework which can "read" a large number of news articles and extract the semantic relatedness between cities. This framework is based on a natural language processing model and employs a machine learning process to identify the main topics of news articles. We describe the overall structure of this framework and its individual modules, and then apply it to an experimental dataset with more than 500,000 news articles covering the top 100 U.S. cities spanning a 10-year period. We perform exploratory visualization of the extracted semantic relatedness under different topics and over multiple years. We also analyze the impact of geographic distance on semantic relatedness and find varied distance decay effects. The proposed framework can be used to support large-scale content analysis in city network research.

Extracting and Analyzing Semantic Relatedness between Cities Using News Articles

1. Introduction

News articles are rich sources of information. They cover a variety of topics about our society, ranging from business, economy, and politics, to arts, sciences, and sports. Many types of entities are also mentioned in news articles, which include not only significant organizations, persons, and events, but also general individuals and everyday activities (Newman et al. 2006, Shahaf and Guestrin 2010). Information in news articles is often timely. To attract attention from potential readers, most journalists and news writers attempt to report the latest events promptly (Reuters 2008, Cochrane 2016). Nowadays, many news articles are published online, and the rich and timely information in these digital articles becomes useful data resource for answering scientific questions (Bingham 2010, Wang and Stewart 2015).

Cities, as centers of social, economic, and general human activities (Friedmann 1986, Sassen 2001, Bettencourt et al. 2007), are frequently mentioned in news articles. Two or more cities may co-occur in the same news article, and such co-occurrences can suggest certain *relatedness* between the mentioned cities. Cities can be related under a variety of topics. A news article may report two cities for their similarity (or dissimilarity) in social and environmental conditions (Liu et al. 2014). Urban actors, such as persons, organizations, and corporations, also play important roles in relating one city to another (Taylor, Catalano and Walker 2002, Hoyler and Watson 2013, Salvini and Fabrikant 2016). For example, a politician living in one city may give a campaign speech at another city; a sports team may travel from its home city to multiple cities during a game season; an advanced producer service (APS) firm or a nongovernmental organization (NGO) based in one city may establish a branch office in another city (Sassen 2001, Toly et al. 2012). There can also be flows between cities in the forms of human migrations (Friedmann 1986), transportation (Miller and Shaw 2001, Smith and Timberlake 2001, Neal 2010), information exchange (Castells 1996, Miller 2004), and many others. As a result, cities are interrelated into a network, in which the nodes are cities and the edges can have different semantics, such as persons, firms, and many types of flow (Taylor 2001, Taylor and Derudder 2016). We can also use a two-mode network to represent these relations if we model the edge semantics (e.g., firms) as nodes in addition to the city nodes (Neal 2008, Neal 2012, Liu and Derudder 2012). By reading news articles, one can grasp the general topics that relate the mentioned cities. Yet, given hundreds of thousands of news articles, it is very difficult, if not impossible, for any individual to manually read and understand these articles (Flaounas et al. 2013).

This paper aims to develop a computational framework that can "read" a large number of news articles and uncover the various relatedness between cities. We adopt the notion *semantic relatedness* from the previous GIScience literature (Hecht and Raubal 2008, Hecht et al. 2012, Ballatore, Bertolotto and Wilson 2014), and use it to refer to the city relatedness under different semantic topics. Semantic relatedness is also studied in the fields of natural language processing (NLP) and computational linguistics where the goal is often to understand the possible concepts or human knowledge that link distinct words together (Budanitsky and Hirst 2006). A good example given by Gabrilovich and Markovitch (2007) is that: although the words *cat* and *mouse* refer to biologically different animals, most people would consider them as related, since there exist films, books, and personal stories that link these two words together. In this work, we focus on the semantic relatedness between cities, and employ news articles as a spatiotemporal data resource for extracting such relatedness information.

There are several advantages in using news articles for extracting semantic relatedness between cities. First, news articles are information sources that can be accessed relatively easily (Taylor 1997, Beaverstock et al. 2000). This feature can help remedy the data deficiency in city network research (Smith and Timberlake 1995, Short et al. 1996). With a news article dataset, we can use the computational framework proposed in this work to study the relatedness of a large number of cities in a timespan of multiple decades. Second, news articles capture diverse city relations. As suggested by Castells (1996) and Alderson and Beckfield (2004), cities can be linked through many relations, such as politics, culture, economy, and technology. News articles, which cover many different topics, seem to be a suitable data source for extracting such diverse city relatedness. Third, the timely nature of news articles enables a temporal exploration of the extracted semantic relatedness based on the publication dates of news articles. Therefore, we can perform longitudinal analyses to understand the changes in inter-city relations over time (Taylor, Catalano and Gane 2003, Taylor and Aranya 2008, Derudder et al. 2010). Finally, news articles can help discover the intangible city relatedness perceived by people. News articles capture not only the socioeconomic activities that happened in our physical world, but also some of the cultures, interests, and public concerns which exist only in the perception of people. Analyzing the contents of news articles, thus, could help us understand such intangible relatedness.

By extracting semantic relatedness between cities, our work has potential applications in a number of areas. In city planning and policy making, the extracted semantic relatedness can reveal the positions of individual cities in the city networks under different semantic topics, and can help identify the current and potential collaborating cities in these networks. City planners and policy makers can then provide conducive policies or construct suitable facilities to improve the positions of their cities in the network (Taylor 2001), and evaluate the effectiveness of their policies by comparing the city networks over multiple years. In geographic information retrieval (GIR) (Jones and Purves 2008), semantic relatedness can be used for searching relevant information based on the input query of a user. Given a query, e.g., "crimes in Chicago", a GIR system would suggest information about cities related to "Chicago" under the topic of "crime", which may not be the same set of cities related to "Chicago" under the topic of, e.g., "sports". Such a feature can be especially useful in geoportals to help users find relevant maps and geospatial data (Hu et al. 2015). In addition, the extracted semantic relatedness can be integrated with existing research on place-based GIS (Goodchild 2011, Gao et al. 2017) to computationally model human cognition on place relations. There also exist other possible applications of semantic relatedness, such as community detection (Expert et al. 2011) and spatial network visualization (Guo 2009, Rae 2009).

The main contributions of this paper are twofold:

- First, we propose a computational framework that can "read" a large number of news articles (more than 500,000 articles in our experiment) and extract semantic relatedness between the mentioned cities.
- Second, we find varied distance decay effects of the semantic relatedness extracted from a news article dataset after analyzing the distances between cities and their relatedness under different topics.

In addition, we perform exploratory visualizations on the multiple city networks that can be derived from the extracted semantic relatedness. We also explore the temporal variations of the semantic relatedness between cities over years. The remainder of this paper is organized as follows. Section 2 reviews the existing work on place name co-occurrence, geographic information retrieval, place relatedness and city networks. Section 3 formalizes the problem of semantic relatedness extraction, and presents the overall structure of our computational framework as well as its individual modules. Section 4 applies the proposed framework to a dataset with more than 500,000 news articles to extract semantic relatedness between cities and then visualizes the result. Section 5 performs a distance decay analysis based on the extracted semantic relatedness and the geographic distances between the cities. Finally, Section 6 summarizes this work and discusses future directions.

## 2. Related work

The co-occurrence of place names has intrigued many GIScience researchers. It is generally assumed that the existence of co-occurrence (often in a textual context) indicates certain relatedness between the mentioned places (Hecht and Raubal 2008, Twaroch, Jones and Abdelmoty 2009, Liu et al. 2014, Ballatore et al. 2014, Spitz, Geiß and Gertz 2016). To extract place name co-occurrences, different types of textual documents can be used, such as Wikipedia articles (Overell and Rüger 2008), news articles (Liu et al. 2014), and general Web pages (Jones et al. 2008). Two place names are considered as "co-occurring" if both are mentioned in a pre-defined textual context. Full documents, such as an entire Wikipedia article or a news article, are often used as the context for extracting co-occurrences (Hecht and Moxley 2009, Liu et al. 2014, Salvini and Fabrikant 2016), although individual paragraphs or sentences can also be used as the contexts. The extracted place name co-occurrences can then be used for various applications and analyses.

One major application of place name co-occurrence is in geographic information retrieval, especially place name disambiguation. Overell and Rüger (2008) analyzed the co-occurrences of place names in Wikipedia articles, and developed a set of rules for disambiguating place names. Based on these rules, if the name *Paris* shows up followed by *Lamar County*, then it should refer to *Paris, Texas* rather than the more prominent *Paris, France*. Also based on the co-occurrences of places in Wikipedia articles, Geiß et al. (2015) and Spitz et al. (2016) constructed toponomy networks for place name disambiguation. A unique feature of their work is the inclusion of textual distances between the mentioned places to quantify the weights of their co-occurrences. In some studies, the co-occurrences of places are combined with the co-occurrences of other types of entities, such as companies and universities, to improve the accuracy of place name disambiguation (Cucerzan 2007, Hu, Janowicz and Prasad 2014, Ju et al. 2016). In addition, place name co-occurrences were also used for identifying the indeterminate boundary of vague place names (Jones et al. 2008, Twaroch and Jones 2010).

Another line of research based on place name co-occurrences examines place relatedness as well as its spatial and temporal implications. Tobler's First Law (TFL) provides a fundamental guidance on place relatedness with regard to geographic distances (Tobler 1970). To empirically validate TFL, Hecht and Moxley (2009) examined the place name co-occurrences in Wikipedia articles written in 22 different languages, and found that the first law holds for the large volume of multilingual Wikipedia data. Based on the place name co-occurrences in news articles, Liu et al. (2014) found that place relatedness in news articles has a weaker distance decay effect compared with those derived from human movements in the physical world. Also using news articles, Zhong et al. (2017) found that places are more likely to co-occur if they are in the same administrative level or have a geographic part-whole relation (e.g., *Los Angeles* is part of *California*). Yuan (2017) explored the temporal variation of place relatedness by examining the place name co-occurrences in a global event dataset extracted from Web news. These previous studies, however, generally use a single value to quantify the relatedness between two places. While valuable and convenient to use, a single value cannot capture the diverse relatedness between places.

In city network research, news articles and newspapers have been considered as a useful information source for studying city relations. Back in 1980, Pred (1980) used the daily newspapers of nine U.S. cities and their mentioning of nonlocal place names to analyze the development of city relations and city-systems between 1851 and 1860. As noted by Pred (1980), performing such content analysis on even a very small sample of newspapers took a lot of efforts. As a result, only three to five newspaper issues were sampled for each city for the analysis. Employing the approach from Pred (1980), Taylor (1997) performed content analysis based on six local newspapers (business sections only) in six major U.S. cities and took a sample of 24 days of the newspapers. This dataset was re-used in a later study by Beaverstock et al. (2000) who considered place mentioning in newspapers as a "surrogate" measure of city relations. Although these previous studies used a small sample of newspaper data, they shed valuable insights on city relatedness and networks. More recently, computational methods were utilized to understand cities and city relations from large text corpus. Chen, Yan and Zhang (2016) analyzed the mentioning frequency of city names in Google Books N-gram corpus and New York Times to understand the changes of international visibility of major Chinese cities between 1700 and 2000. While Chen et al. (2016) focused on individual cities, Salvini and Fabrikant (2016) examined city relatedness using their co-occurrences in Wikipedia articles. A notable feature of Salvini and Fabrikant (2016)'s work is that they utilized not only the co-occurrence counts, but also the article categories (which were assigned by Wikipedia users) to semantically annotate the relatedness. The authors then constructed multiple city networks under the different functions, such as "Politics" and "Economy/technology", derived from the user-generated categories. Our work aligns with the general direction of Salvini and Fabrikant (2016), but also has several unique characteristics. One is our use of news articles instead of Wikipedia articles, which enables longitudinal analysis of the extracted semantic relatedness based on the news publication dates. In addition, the semantic topics in our work are identified by analyzing the full texts of news articles using a natural language processing (NLP) model rather than relying on only the categories or tags assigned by users, which can be incomplete. Finally, our work examines the varied distance decay effects of semantic relatedness between cities, which, to our best knowledge, were not studied before.

If we go beyond the literature of GIScience and urban geography, place name co-occurrence can be considered as a special case of word co-occurrence studied in information retrieval, natural language processing, and computational linguistics (Deerwester et al. 1990, Morris and Hirst 2004). Wikipedia articles were also frequently used in these domains to study the semantic relatedness between words, such as *WikiRelate!* from Strube and Ponzetto (2006) and the Explicit Semantic Analysis proposed by Gabrilovich and Markovitch (2007). Before the existence of Wikipedia, general lexical resources, such as dictionaries, were used for computing semantic relatedness. For example, Kozima and Furugori (1993) converted the Longman Dictionary of Contemporary English into a network by linking each headword with the words in its definition, and semantic relatedness was represented by the paths between two words. WordNet (Miller 1995) is another lexical resource that was widely used for semantic relatedness studies (Pedersen, Patwardhan and Michelizzi 2004, Patwardhan and Pedersen 2006, Taieb, Aouicha and Hamadou 2014). Compared with these previous studies, our work has a focus on geographic space, and examines the semantic relatedness between cities.

## 3. Computational framework

### 3.1 Problem definition

Before presenting the details of the framework, we first formalize the problem of semantic relatedness extraction. Let a set $C$, with elements $c_1, c_2, \ldots c_n$, represent the cities whose semantic relatedness we would like to study, and let a set $T$, with elements $t_1, t_2, \ldots, t_m$, represent the time periods of interest. Based on $C$ and $T$, we can collect a set of news articles $D$ which contain the co-occurrences of the cities in $C$ and which were published during the time periods of $T$. Let a set $Y$, with elements $y_1, y_2, \ldots, y_l$, represent the semantic topics of our interest. Depending on the application requirements, $Y$ can be defined as a set of topics covering different domains, such as "culture", "environment", "politics", and "education", or $Y$ can be defined to focus on one specific domain, such as "education", with detailed topics, such as "higher education" and "K-12 education". Using the news articles in $D$ and the defined topics in $Y$ as the input, we hope to obtain an output $R$ which contains the quantitative relatedness between cities under different topics in multiple years. Figure 1 illustrates the formalized input and output of the problem of semantic relatedness extraction.

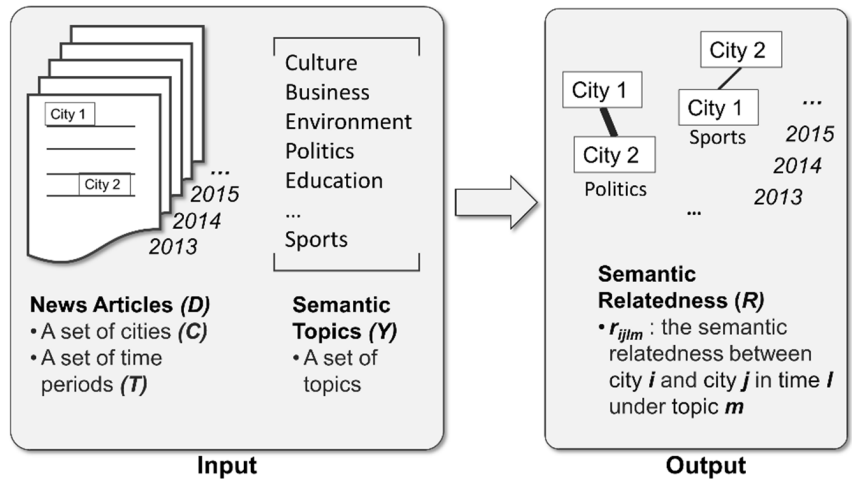

Figure 1. The input and output of the formalized problem of semantic relatedness extraction.

*3.2 The framework*

The computational framework is proposed to help us derive the output from the input. The overall framework is a machine learning process. The general idea is to train a natural language processing model, use this model to "read" news articles and extract their main topics, and then annotate the semantics of city relatedness based on the topics of the news articles. The strength of the semantic relatedness between two cities under one topic in a time period is quantified as the number of news articles that contain the city co-occurrences under this topic and were published during this time period. Figure 2 provides an overview of the framework.

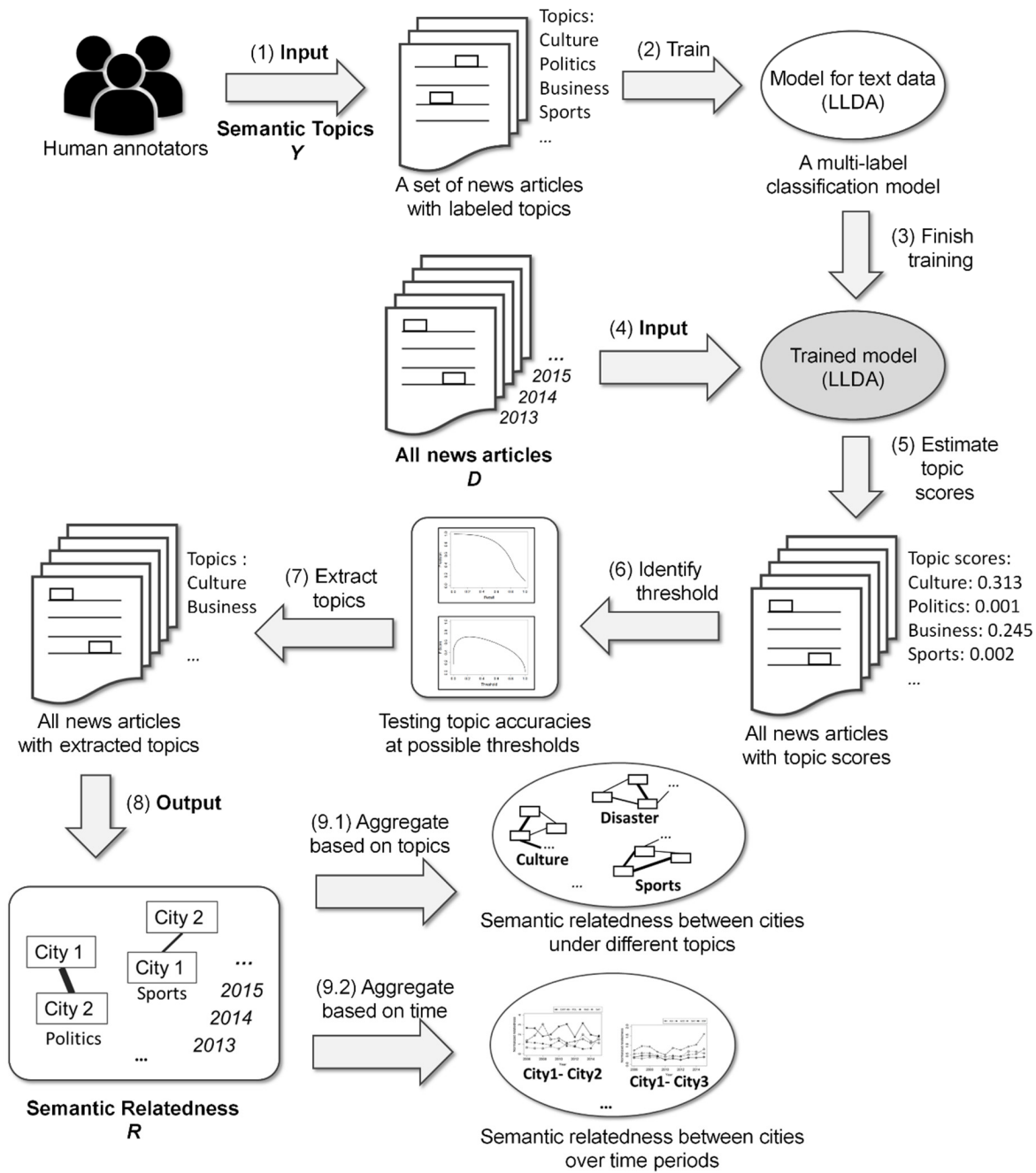

Figure 2. The overall structure of the proposed computational framework.

The framework is based on the following steps. In step (1), we develop a training dataset using news articles and the semantic topics of interest. Like many other machine learning problems (Marcus, Marcinkiewicz and Santorini 1993, Nivre et al. 2016), a training dataset is generally necessary for machines to learn human knowledge. Here, machines are expected to learn the knowledge of identifying the main topics discussed by a news article. Accordingly, we need a training dataset, in which news articles are annotated with the corresponding semantic topics defined in $Y$. Such a training dataset can be obtained with the help of human annotators. Crowdsourcing platforms, such as Amazon Mechanical Turk (Paolacci, Chandler and Ipeirotis 2010), can also be employed to collect the training dataset. Sometimes, we can utilize the special features of a particular dataset to develop the training dataset in a relatively fast manner, and we will demonstrate such a possible way in the later experiment section.

Step (2) uses the obtained training dataset to train a multi-label text classification model. A multi-label classification model is employed since one news article (e.g., an article about "climate change") can discuss multiple topics (e.g., both "environment" and "politics"). Specifically, we use the Labeled Latent Dirichlet Allocation (LLDA) model (Ramage et al. 2009), which is a generative model that considers a textual document as being generated from a mixture of multiple topics. We use the LLDA model in the proposed framework for two major reasons. First, the LLDA model is a supervised natural language processing model which allows the extracted topics to be confined to a set of target topics defined in $Y$. This can be differentiated from the more commonly used unsupervised Latent Dirichlet Allocation (LDA) model, which may extract some topics that are abstract and hard to interpret (Blei, Ng and Jordan 2003). Second, previous research has shown that the LLDA model has better performance in multi-label classification when compared with other supervised models, such as support vector machines (Ramage et al. 2009) and Naïve Bayesian models (Hu et al. 2015). In the training process, the LLDA model examines the word frequencies in a news article and its corresponding topics in the training dataset, and then gradually learns the associations between the semantic topics and the word probabilities.

Step (3) completes the training process of the LLDA model, and in step (4), all news articles mentioning the cities in $C$ and published during the periods in $T$ are fed to the trained LLDA model. In step (5), the trained LLDA model looks into the full texts of these news articles, and estimates the scores of different topics for each news article. A topic score estimated by the LLDA model is in [0, 1], and the sum of the topic scores is 1. In step (6), we develop an automated process for identifying a suitable threshold for extracting the topics. Such a process iterates the possible threshold values from 0 to 1, and identifies the suitable threshold value at which the LLDA model achieves its best performance. Details about this automatic process will be further demonstrated in the following experiment section with specific data records. In step (7), the topics whose scores are higher than the identified threshold are extracted as the topics of the news articles, and multiple topics can be extracted from one news article. In step (8), we quantify the semantic relatedness between cities based on the numbers of news articles under different topics. This step achieves our expected output of extracting semantic relatedness between cities under different topics in multiple time periods.

The extracted semantic relatedness can be aggregated in different ways to facilitate the result exploration. In step (9.1), we can aggregate the semantic relatedness based on topics, such as "Culture" and "Disaster". Accordingly, multiple city networks can be constructed, with each network under a distinct topic. In step (9.2), we can aggregate the semantic relatedness of particular city pairs based on time periods (e.g., years). Accordingly, we can examine the fluctuations of the semantic relatedness between cities over multiple years. In the following section, we will apply the proposed computational framework to a news article dataset to extract semantic relatedness between cities.

## 4. Experiment

*4.1 Dataset*

In this experiment, we examine the semantic relatedness among the top 100 cities (ranked by population) in the contiguous U.S. from January 1, 2006 to December 31, 2015. These major cities are selected because they are often mentioned in news articles and therefore have rich materials for examining their semantic relatedness. If we relate back to the problem definition, $C$ is the 100 cities in Figure 3, and $T$ is the 10 years from the beginning of 2006 to the end of 2015.

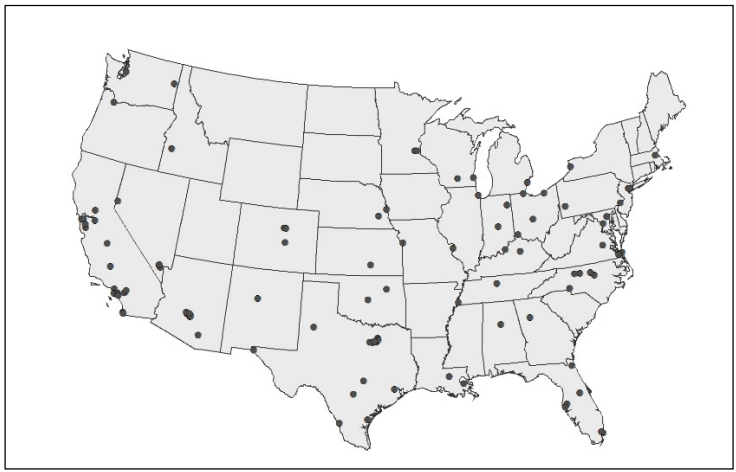

Figure 3. Locations of the top 100 cities in the contiguous U.S.

The news articles for this experiment were retrieved from The Guardian, which is a daily newspaper that publishes news articles covering the entire world (https://www.theguardian.com). We chose The Guardian as the data source of our experiment for three major reasons. First, The Guardian is a general news outlet, which covers a wide range of news topics rather than narrowly focusing on one or several topics. These various topics allow us to discover the diverse semantic relatedness between cities. Second, The Guardian is a well-recognized news source. As a newspaper with almost 200-year history, The Guardian has a good reputation and is unlikely to publish fake news. Third, The Guardian provides an open Application Programming Interface (API) which offers access to a rich amount of news data. With the slogan "Open to everyone" and "Access over 1.9 million pieces of content" (http://open-platform.theguardian.com/), the API allows users to retrieve not only the metadata of news (e.g., titles, publication dates, and news sections) but also the full texts of all the news articles for each city pair in our study. The data access of other news outlets is more restricted, such as *The Wall Street Journal* (which does not provide an API) and *The New York Times* (which offers an API to the metadata but not to the full texts; it also limits the number of the returned records to 1,000 per city pair). Nevertheless, using the news articles from one news outlet may introduce bias to the experimental result. In this work, The Guardian data are used for demonstrating the general effectiveness of the proposed computational framework, and our framework is not restricted to the news articles from specific outlets. When more news article data (especially full text data) become openly available, we can apply the framework to other datasets and compare the results.

We systematically retrieved the news articles from The Guardian using its API. The entire dataset consists of two parts: 1) the news articles that contain the co-occurrences of these cities; and 2) the news articles that contain individual cities. For the first part, we iterated through each city pair (4,950 city pairs in total) based on the 100 cities, and retrieved the news articles that contain both cities. 265,436 articles were retrieved (1,105 city pairs are not found to occur in any news article). For the second part, we iterated through the 100 cities, and retrieved the news articles that contain each individual city using the API. In total, 278,388 news articles were retrieved (each of the 100 cities is mentioned by news articles; the least number of news articles for one city is 5). The second part of the dataset was collected for performing the distance decay analysis in Section 5. Combining these two parts, we retrieved more than 500,000 news articles for this experiment.

*4.2 Performing the experiment*

We first define the semantic topics $Y$ for studying the relatedness between cities. As described previously, these topics should be defined based on the application requirements. In this experiment, we employ the standard news topics from the International Press Telecommunications Council (IPTC) (https://iptc.org/) which provides a hierarchy of topics about various aspects of our society. Specifically, the 17 general topics at the first level (down from the root) of the IPTC topic hierarchy are used, and a complete list of these topics is provided in the left column of Table 1.

Table 1. The 17 general IPTC topics and their corresponding news tags.

| IPTC Topic | News Tags |
| --- | --- |
| Arts, Culture and Entertainment | culture, music, film, media, books, artanddesign, television, art, fashion, festivals, history, comedy, museums, opera, drama, poetry, documentary, painting, theatre, sculpture |
| Crime, Law and Justice | crime, law, rape, supreme-court, human-rights, police, criminal-justice |
| Disaster, Accident and Emergency Incident | natural-disasters, hurricane, flooding, earthquake, drought, wildfire |
| Economy, Business and Finance | business, economics, banking, advertising, transport, economy, market, realestate, investing |
| Education | education, schools, teaching, students |
| Environment | environment, climate-change, energy, water, pollution, waste |
| Health | health, healthcare |
| Human Interest | awards-and-prizes, celebrity, animals |
| Labor | labour, employment, unemployment |
| Lifestyle and Leisure | lifeandstyle, travel, food-and-drink, hotels, restaurants, bars |
| Politics | politics, democrats, republicans, election, policy |
| Religion and Belief | religion |
| Science and Technology | technology, internet, research, science, biology, psychology, software, genetics, mathematics, chemistry |
| Society | society, race, communities, poverty, family, homelessness, immigration, marriage, population, migration |
| Sport | sport, nfl, nba, football, basketball, baseball, boxing, tennis, cricket, olympics, athletics, swimming, cycling |
| Conflicts, War and Peace | protest, terrorism, military |
| Weather | weather |

To develop the training dataset, we make use of the tags associated with the news articles. These tags were assigned by news producers (Guardian 2014), and one news article can be associated with from zero to many tags. The tags have varied granularities, ranging from general topics, such as "culture" and "business", to specific entities and events, such as "michael-jackson" and "world-aquatics-championships-2011" (in total 14,376 distinct tags are found in the dataset). We develop the training dataset by building a topic-tag linking table. The authors read through the tags associated with at least 100 news articles, and identify the tags that can be obviously linked to the IPTC topics. Generally, links are established when an IPTC topic (e.g., "Arts, Culture and Entertainment") explicitly contains a tag (e.g., "culture"), or when a tag (e.g., "hurricane") can be considered as a hyponym of an IPTC topic (e.g., "Disaster, Accident and Emergency Incident"). When disagreement happens, we discard the tags and only keep those whose IPTC topics are agreed. The topic-tag linking table is shown in Table 1. It is worth noting that other human annotators might produce a topic-tag linking table different from Table 1. Involving domain experts can help ensure that the topic-tag links are generally agreed within the application area. News articles with the tags found in Table 1 are assigned to the corresponding IPTC topics, and are added into the training dataset. In total, we obtained 141,765 training data records (each data record contains one news article and one or more IPTC topics). The number of training data records varies across the topics. For example, there are

20,605 data records under the topic "Politics", 7,075 data records under the topic "Environment", and only 917 records under the topic "Weather" (which is the least number of records under a topic). Generally, more data records can provide better training for a topic.

We train the LLDA model using the developed training dataset. In the training process, the model learns the probabilistic associations between the IPTC topics and the words. Figure 4 shows the top 20 most frequent words learned for three IPTC topics: "Politics", "Environment", and "Weather". Based on these words, it seems that the LLDA model has learned reasonably well about these topics, including the topic "Weather" which, as mentioned before, has relatively fewer number of training data records compared with the other two topics.

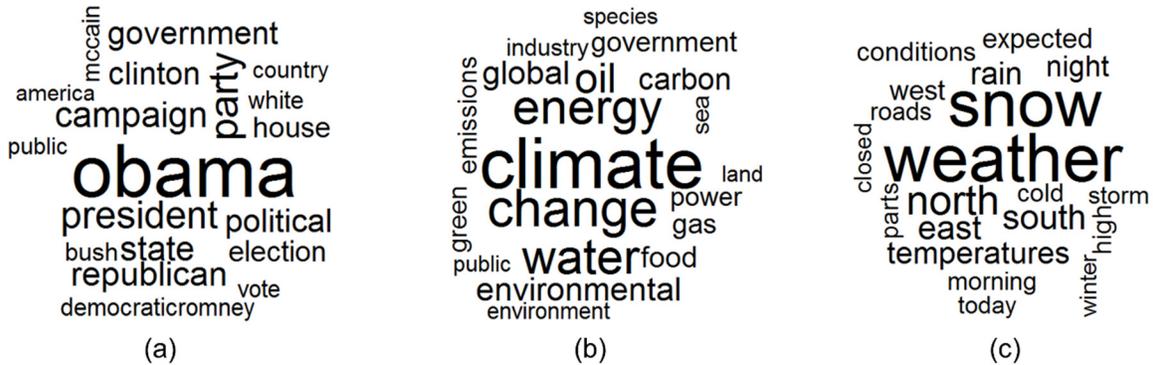

Figure 4. Three IPTC topics and their top 20 most frequent words learned by the LLDA model: (a) Politics; (b) Environment; and (c) Weather.

We apply the trained LLDA model to all news articles in our experimental dataset to extract topics. Based on the learned knowledge about the 17 IPTC topics and the observed word frequencies in the news articles, the LLDA model estimates 17 probability values (topic scores) for each news article to quantify its likelihood of discussing the 17 IPTC topics respectively.

The next step is to determine a suitable threshold value so that topics with scores higher than the threshold can be extracted. Here, the goal is to identify a threshold based on which the extracted topics are both relevant and complete. Two metrics, *precision* and *recall*, from information retrieval are employed to quantify the relevance and the completeness respectively. Precision (Equation 1) measures the percentage of the extracted relevant topics among all extracted topics. Recall (Equation 2) measures the percentage of the extracted relevant topics among all relevant topics.

$$Precision = \frac{|Extracted\ Relevant\ Topics|}{|All\ Extracted\ Topics|} \qquad (1)$$

$$Recall = \frac{|Extracted\ Relevant\ Topics|}{|All\ Relevant\ Topics|} \qquad (2)$$

Precision and recall are often a trade-off. Generally, a high threshold produces a high precision but a low recall. This is because only the topics with very high scores are extracted, and some relevant topics with scores lower than the defined high threshold are excluded. In contrast, a low threshold generally produces a high recall but a low precision. To achieve a balance between the two, we employ a third metric, F-score, which is the harmonic mean of precision and recall (Equation 3). F-score will have a high value only when both precision and recall are fairly high, and will have a low value when either of the two is low.

$$F\text{-}score = 2 \cdot \frac{Precision \cdot Recall}{Precision + Recall} \qquad (3)$$

We identify the most suitable threshold using the training dataset in which the relevant topics of news articles are known. A 10-fold cross-validation (Flach 2012) is employed to identify a robust threshold. In this process, the entire training dataset is divided into 10 folds, and each time 9 folds of data are used for training a LLDA model and 1 fold of data is held out for testing the F-score. We iterate the threshold from 0 to 1 with an incremental step 0.001; at each threshold, we train and test the LLDA model 10 times to ensure each fold of data has been used for both training and testing; and the final F-score at a threshold is the average of the obtained 10 F-scores. This 10-fold cross-validation process can help increase the robustness of the derived threshold and avoid overfitting (Flach 2012). Figure 5 shows the threshold values from 0 to 1 in the $x$ axis and the corresponding average F-scores in the $y$ axis. At the threshold 0.227, the F-score achieves its maximum 0.71 with a precision 0.76 and a recall 0.66. Thus, we use 0.227 as the threshold, and topics whose scores are higher than 0.227 are extracted as the topics of the news articles.

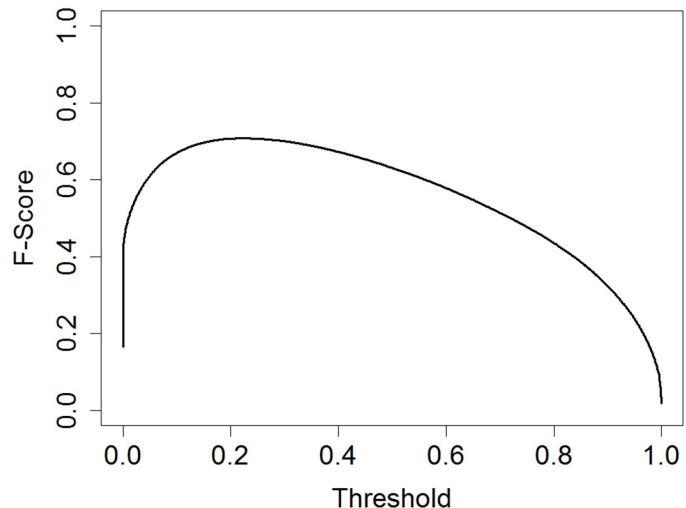

Figure 5. Threshold values in [0, 1] and the corresponding F-scores.

We quantify the semantic relatedness between cities based on the topics extracted from the news articles. The semantic relatedness between two cities under a topic is quantified as the number of news articles which contain the co-occurrences of the two cities and which also discuss this topic. We can further quantify the semantic relatedness between two cities under a topic and in a specific year using the number of news articles published in that year.

*4.3 Results and discussion*

The direct output of our experiment is a 4,950-by-17 matrix with each row representing one city pair, each column representing one IPTC topic, and each cell containing the relatedness value of a city pair under a topic. By distinguishing the semantic relatedness by years, we obtain 10 matrices, each of which is 4,950 by 17 and contains the semantic relatedness in each year. We can visualize the extracted semantic relatedness to facilitate the exploration of the result. In the following, we provide two exploratory visualizations based on the topics and the years respectively.

Based on the topics, we can aggregate the semantic relatedness into multiple city networks with each under one topic. Specifically, we can construct 17 city networks based on the 17 IPTC topics. In one network, nodes represent cities and edges represent the relatedness between cities. To construct these networks, we first derive 17 city-to-city matrices from the extracted semantic relatedness. Each matrix is 100 by 100. The values in a matrix represent the relatedness between the city of the row and the city of the column under one IPTC topic. While the city-to-city matrices can already be displayed as networks, our initial visualization showed a lot of clustered nodes and edges. This is because these networks are rather complex and one city can be linked to many other cities. To simplify these networks while still preserving the fundamental network structures, we use the Pathfinder network scaling (PFNet) algorithm (with $q = n - 1, r = \infty$) (Schvaneveldt 1990), which is an effective network pruning method often used in information visualization (Skupin 2014, Salvini and Fabrikant 2016). The output of the PFNet algorithm is added to a network visualization software Gephi, and we employ its default layout algorithm, ForceAtlas2

(Jacomy et al. 2014), to configure the layout of the nodes and edges. Figure 6 shows 4 of the 17 constructed city networks. In each sub figure, the sizes of the nodes represent the weighted degrees of the cities in the current network, and the widths of the edges represent the strengths of the relatedness between the cities. To ensure legibility, we only show the top 30 cities with the highest weighted degrees in each sub figure.

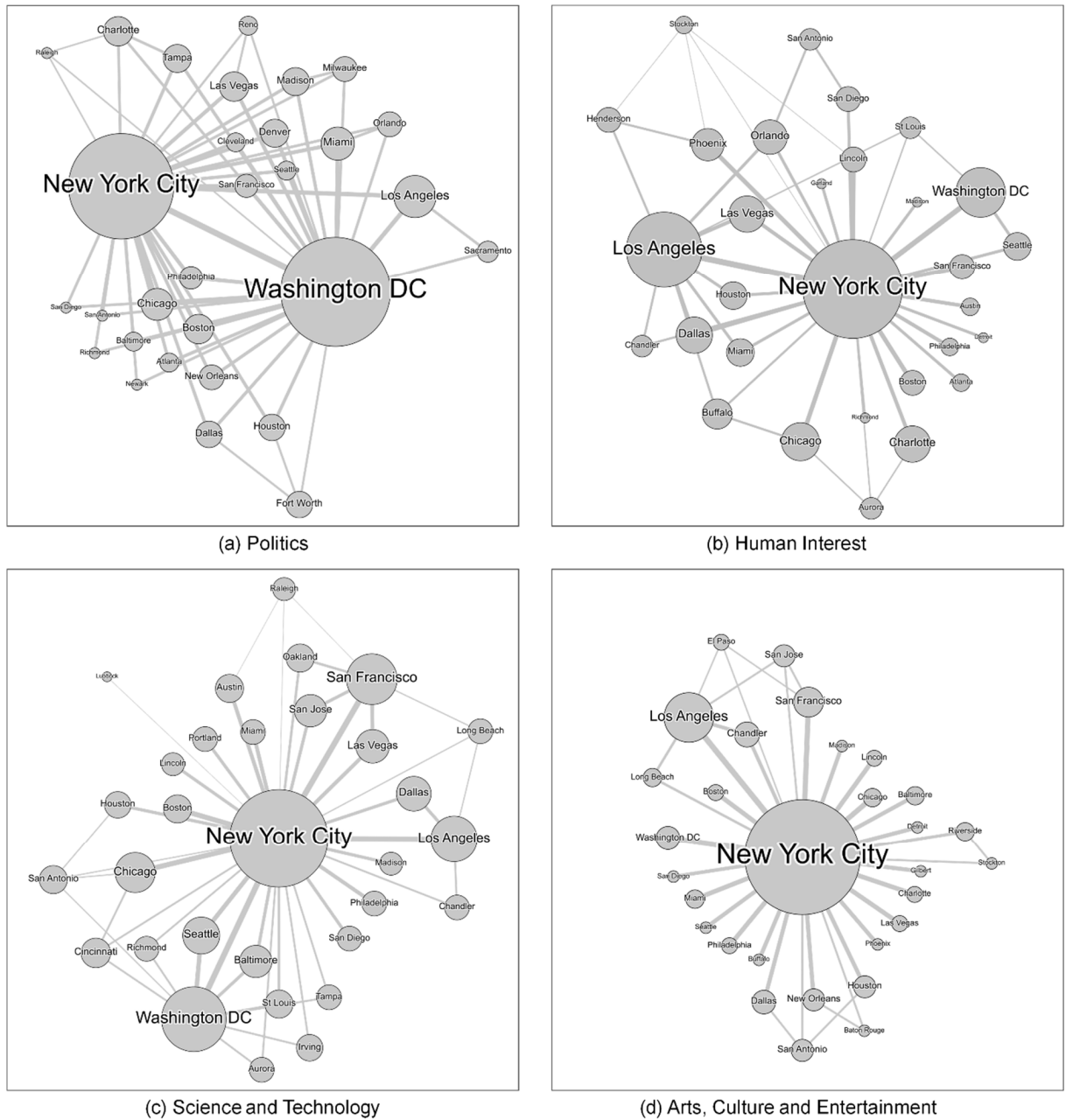

Figure 6. Network visualizations of the semantic relatedness under 4 topics: (a) Politics; (b) Human Interest; (c) Science and Technology; (d) Arts, Culture and Entertainment

It can be seen from Figure 6 that New York City (NYC) plays a dominating role in all four sub figures. While NYC is the most populous city in the United States, which indeed has prominent impacts on politics, business, culture, and many other fields (Friedmann 1986, Sassen 2001), this result might also be attributed to the location bias of the experimental dataset from The Guardian whose US main office is in NYC. The other cities in the four networks show varied significances under different topics. For example, Washington DC plays a much more significant role in the city network under the topic "Politics" compared with its roles in the other three networks. Los Angeles, the home of Hollywood, is more prominent in the network under the topic "Human Interest", which covers a lot of news about celebrities and awards (e.g., the Oscars), than in the network under the topic "Politics". A similar situation also applies to San Francisco, which is an important hub in the network under the topic "Science and Technology", but plays less significant roles in the other three networks. Under the topic "Arts, Culture and Entertainment", we see a more centralized role of NYC compared with the other three networks which show more of a multi-center structure. Based on these city networks, we can also infer that the cities which function as major hubs in multiple networks should be more diverse than those that are hubs in only one or two networks.

Based on the publication dates of news articles, we can visualize the extracted semantic relatedness as line charts. Specifically, we can create a line chart for each city pair to illustrate the temporal variation of their semantic relatedness over 10 years. To construct these line charts, we first normalize the semantic relatedness values using Equation 4:

$$n_{ijl} = \frac{c_{ijl}}{\sum_l \sum_i \sum_j c_{ijl}} \times 100 \tag{4}$$

where $n_{ijl}$ represents the normalized relatedness between city $i$ and city $j$ in year $l$, $c_{ijl}$ is the original count of news articles, and $\sum_l \sum_i \sum_j c_{ijl}$ is the sum of all news articles among all city pairs and years. The multiplication factor 100 is used for reducing underflow, since $\frac{c_{ijl}}{\sum_l \sum_i \sum_j c_{ijl}}$ can be a very small number. We perform this normalization because different topics may receive different amounts of media attention in general. For example, there are more news articles about "Sport" than "Religion and Belief" in our experimental dataset. Thus, the normalization allows the relatedness under different topics to be plotted in the same chart for a fair comparison. Figure 7 provides the line chart visualizations of the semantic relatedness of 4 city pairs. For legibility, we only show the top 4 topics that have the strongest normalized relatedness in each sub figure.

It can be seen in Figure 7 that the semantic relatedness between two cities do not remain the same. Instead, they fluctuate over the years and sometimes the variations can be large. For example, in Figure 7(a), NYC and Washington DC had a much stronger relatedness under the topic "Politics" in 2008 than in other years, probably due to the U.S. presidential election in which the U.S. had its first African American president. In Figure 7(d), Los Angeles and New Orleans had a peak relatedness under the topic "Disaster, Accident and Emergency Incident" in 2014. To understand the reason, we read some news articles about these two cities in this year in our dataset, and find multiple articles about the "100 Resilient Cities" program, which was designed to help cities become more resilient to disasters. This program started in Dec. 2013 and received a lot of applications from cities in 2014 (http://www.100resilientcities.org/about-us). Both Los Angeles and New Orleans were enrolled in this program. Figure 7 also shows that each city pair has its uniqueness in terms of the main topics of their semantic relatedness. The top topics between NYC and Washington DC are "Conflicts, War, and Peace" (which includes news about military actions and protests) and "Politics". In contrast, Washington DC does not seem to be engaged under the topic of "Politics" with San Francisco, and the main topics between the two cities are "Science and Technology", "Environment", and "Society". The top relatedness topics between NYC and Los Angeles are "Human Interest" and "Arts, Culture and Entertainment", whereas the main relatedness between Los Angeles and New Orleans seems to be a combination of "Arts, Culture and Entertainment" and "Disaster, Accident, and Emergency Incident", reflecting the characteristics of both cities.

The extracted semantic relatedness also opens the door to many other research questions, which can be grouped from spatial, temporal, and thematic perspectives. From a spatial perspective, we can ask questions such as:

- *What is the impact of geographic distance on the semantic relatedness?*

- *What are the location-related factors that may have facilitated the relatedness between cities?*

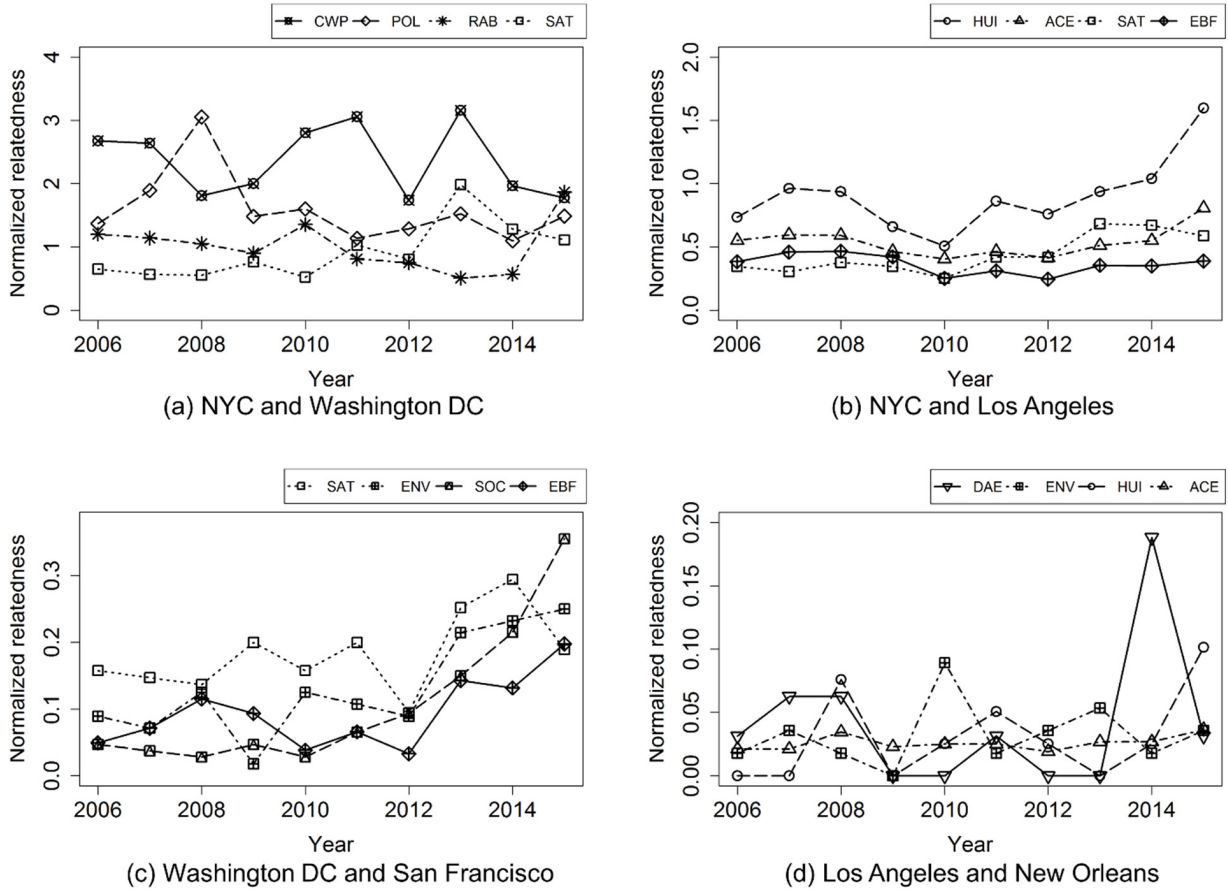

Figure 7. Temporal variations of the semantic relatedness of 4 city pairs. Topics are written in acronyms to fit the figure, which are listed as follows. CWP: Conflicts, War, and Peace; POL: Politics; RAB: Religion and Belief; SAT: Science and Technology; HUI: Human Interest; ACE: Arts, Culture and Entertainment; EBF: Economy, Business and Finance; ENV: Environment; SOC: Society; DAE: Disaster, Accident, and Emergency Incident.

From a temporal perspective, we can ask questions such as:
- *How does the semantic relatedness between cities fluctuate over seasons?*
- *What are some possible reasons for a large change of semantic relatedness over years?*

From a thematic perspective, we can ask questions such as:
- *Which cities are major hubs in more than x number of city networks under different topics?*
- *What are the city communities in the city network under one particular topic?*

While we may not be able to discuss all these questions in this paper, we examine one question in the following section, namely the impact of geographic distance on the extracted semantic relatedness.

## 5. Distance decay analysis

The distance decay effect of place relatedness derived from news articles was investigated previously by Liu et al. (2014). In that study, the authors extracted place relatedness based on their co-occurrences in news articles retrieved from a news outlet, and examined the distance decay effect of the relatedness using a gravity model. They found a weak distance decay effect quantified by a value of 0.2 for the friction factor in the gravity model. However, this previous study analyzed only the general relatedness between places without differentiating the relatedness under different topics. In this work, we investigate the varied distance decay effects of semantic relatedness. We use the previous work from Liu et al. (2014) as a baseline to which our analysis result will be compared.

We employ the same gravity model as used by Liu et al. (2014), as shown in Equation 5:

$$c_{ij} \propto \frac{c_i c_j}{d_{ij}^{\beta}} \quad (5)$$

where $c_{ij}$ is the count of news articles that contain the co-occurrences of cities $i$ and $j$ (city pairs that do not co-occur in any news articles are excluded). $c_i$ and $c_j$ are the counts of news articles that mention these two cities, which were derived from the second part of our dataset. $d_{ij}$ is the geographic distance between the two cities. We use the great circle distance for $d_{ij}$, taking into account the curved surface of the Earth. $\beta$ is the friction factor, and a larger $\beta$ indicates a stronger distance decay effect.

Estimating the value of $\beta$ is a key step in quantifying the distance decay effect. In the previous work, Liu et al. (2014) took a computational approach in which $\beta$ was iterated from 0 to 1 with a step of 0.1. To compare with their work, we take a similar approach but with a finer step of 0.01. For each $\beta$, we calculate the $\frac{c_i c_j}{d_{ij}^{\beta}}$ for all city pairs (4,950 city pairs), perform regression with $c_{ij}$ using $c_{ij} = b_0 + b_1 \frac{c_i c_j}{d_{ij}^{\beta}}$, and obtain the R-squared value. We then plot the $\beta$s and its R-squared values in Figure 8(a). When $\beta = 0.17$, a maximum R-squared value 0.806 is achieved, and Figure 8(b) plots out the corresponding $c_{ij}$ and $\frac{c_i c_j}{d_{ij}^{\beta}}$ values, as well as the fitted line.

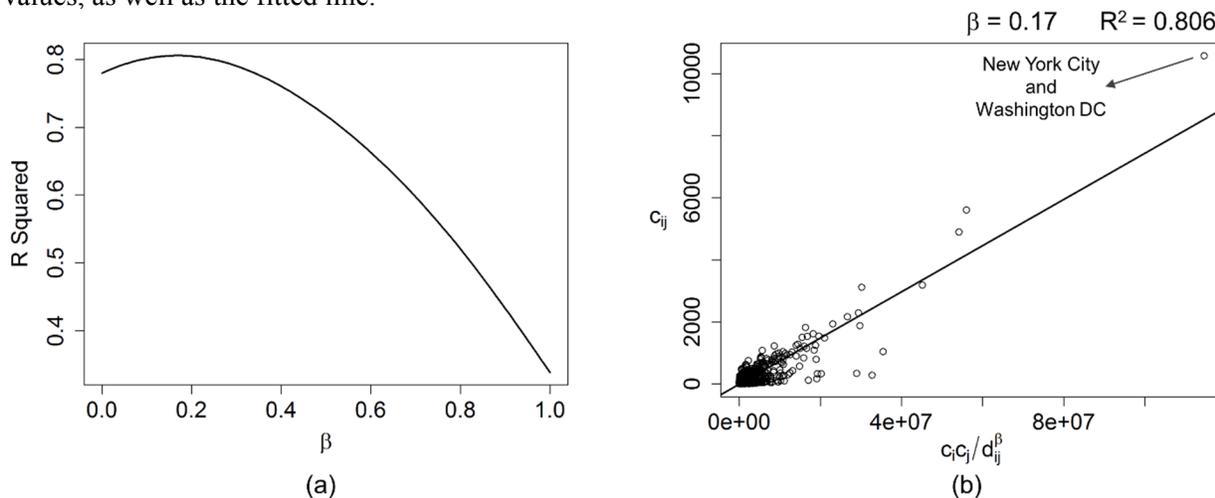

Figure 8. Identifying the friction factor of the gravity model using the counts of news articles: (a) R-squared values at different $\beta$s; (b) the fitted line based on $c_{ij}$ and $\frac{c_i c_j}{d_{ij}^{\beta}}$ when $\beta = 0.17$.

While a R-squared value 0.806 indicates a good fit of the gravity model, this result might be subject to overestimation since most points cluster at the lower left corner of Figure 8(b) (Taaffe, Gauthier and O'Kelly 1996). To obtain a more robust estimate, we apply a log transformation on both $c_{ij}$ and $\frac{c_i c_j}{d_{ij}^{\beta}}$, and re-run the computational experiment. Figure 9 shows the new result based on the log-transformed values.

While the R-squared value has decreased to 0.654, this result is more robust and the obtained best friction factor is 0.23. This value is very close to the 0.2 identified by Liu et al. (2014) using news articles from a different outlet. In fact, the results could have been even closer, since Liu et al. (2014) used 0.1 rather than 0.01 as the iteration step (thus, a value like 0.23 was not tested). This result difference suggests a limitation of the computational approach in which a suitable iteration step needs to be specified. A large step carries the risk of omitting possibly more accurate values, while a small iteration step brings a high computational cost.

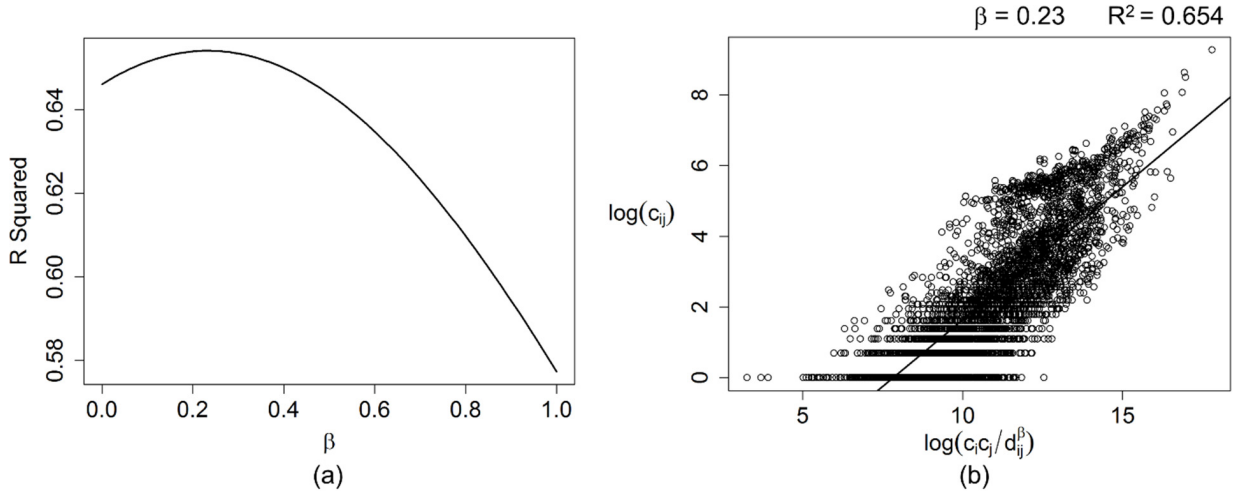

Figure 9. Identifying the friction factor of the gravity model using log-transformed values: (a) R-squared values at different $\beta$s; (b) the fitted line based on $\log(c_{ij})$ and $\log(\frac{c_i c_j}{d_{ij}^\beta})$ when $\beta = 0.23$.

A second examination of the gravity model shows that a suitable $\beta$ can also be obtained through a multiple regression approach, thus avoiding defining the length of the iteration step. In the equation $\log c_{ij} = b_0 + b_1 \log \frac{c_i c_j}{d_{ij}^\beta}$, we can decompose $\log \frac{c_i c_j}{d_{ij}^\beta}$ into $\log c_i c_j - \beta \log d_{ij}$, and obtain $\log c_{ij} = b_0 + b_1 \log c_i c_j + (-\beta * b_1) \log d_{ij}$. A multiple regression can then be performed based on Equation 6:

$$\log c_{ij} = b_0 + b_1 \log c_i c_j + b_2 \log d_{ij} \qquad (6)$$

The coefficient $b_2$ in Equation 6 quantifies the impact of $\log d_{ij}$ on $\log c_{ij}$, and can be directly used to examine the distance decay effect (Taaffe et al. 1996). To compare with the previous work, we calculate $\beta$ using $-b_2/b_1$. We perform a second set of experiments by applying both the computational approach and the multiple regression approach to the news article dataset. These two approaches are first applied to all news articles and then to the news articles under each individual topic. Our objectives are: 1) to compare the results of the two approaches (which should generate similar results); 2) to examine the distance decay effects of the semantic relatedness under different topics. Table 2 shows the result of our experiment.

As shown in Table 2, the results obtained using the two approaches are highly consistent. This result indicates that both approaches are valid, but the multiple regression approach can avoid the difficulty of defining a suitable iteration step. However, in situations when the precision of the result is clearly expected, the computational approach can also be used.

Table 2 also shows varied distance decay effects of the relatedness under different topics. This can be seen from the friction factors derived from the analysis. We can divide the results into two groups based on their R-squared values. For the topics with R-squared values larger than 0.5, gravity models provide fair explanations on the semantic relatedness between the studied cities. Compared with the $\beta$ value 0.23 derived from all news, the semantic relatedness under the topics, such as "Crime, Law and Justice", show a stronger distance decay effect, suggesting that nearby cities are more likely to be discussed together under these topics than cities far apart. In contrast, topics, such as "Sport", show weaker distance decay effects suggesting that news about sports can link almost any cities together regardless of their geographic distances. For the topics with R-squared values smaller than 0.5, the gravity models (in which only $c_i$, $c_j$, and $d_{ij}$ are the explanatory variables) may not provide a good enough explanation, and some additional variables might be included. Nevertheless, two topics are still interesting enough for discussion. First, a strongest friction factor is observed under the topic "Disaster, Accident and Emergency Incident" with a value 0.61, suggesting that nearby cities are more likely to be related under the topic of disasters than cities farther away. This is generally consistent with the impact of most disasters, such as hurricanes, earthquakes, and wildfires, which show a strong spatial effect. Second, the weakest friction factor 0.03 is observed under the topic "Human Interest". This result indicates that human interests can link any cities together even if these cities are far away. In sum, we observed varied distance decay effects in the extracted semantic relatedness between cities, although the relatedness based on all news articles shows a distance decay effect similar to that from the previous study.

Table 2. $\beta$ and $R^2$ obtained for all news (without differentiating the topics) and semantic relatedness (under different topics) using the computational and the multiple regression approaches.

|  | Computational Approach | | Multiple Regression | |
| --- | --- | --- | --- | --- |
| Topic | $\beta$ | $R^2$ | $\beta$ | $R^2$ |
| **All News** | **0.23** | 0.654106 | **0.233219** | 0.654107 |
| Arts, Culture and Entertainment | 0.21 | 0.688115 | 0.211980 | 0.688116 |
| Lifestyle and Leisure | 0.24 | 0.501075 | 0.236083 | 0.501077 |
| **Sport** | **0.08** | 0.509305 | **0.075001** | 0.509308 |
| Environment | 0.43 | 0.461709 | 0.425431 | 0.461713 |
| Economy, Business and Finance | 0.23 | 0.519645 | 0.233249 | 0.519647 |
| Society | 0.29 | 0.483899 | 0.292341 | 0.483900 |
| **Crime, Law and Justice** | **0.37** | 0.549941 | **0.367762** | 0.549942 |
| Science and Technology | 0.19 | 0.569932 | 0.185918 | 0.569935 |
| Politics | 0.32 | 0.599010 | 0.318968 | 0.599010 |
| **Disaster, Accident and Emergency Incident** | **0.61** | 0.482975 | **0.609680** | 0.482975 |
| Health | 0.34 | 0.426866 | 0.339127 | 0.426866 |
| Conflicts, War and Peace | 0.19 | 0.498410 | 0.187540 | 0.498411 |
| Education | 0.16 | 0.431484 | 0.162067 | 0.431484 |
| **Human Interest** | **0.03** | 0.448994 | **0.034326** | 0.448996 |
| Weather | 0.34 | 0.407078 | 0.337029 | 0.407079 |
| Religion and Belief | 0.24 | 0.458105 | 0.236189 | 0.458108 |
| Labor | 0.27 | 0.279421 | 0.267960 | 0.279421 |

## 6. Conclusions and future work

This paper presents a computational framework for extracting semantic relatedness between cities from news articles. News articles cover not only the socioeconomic activities that happened in the physical world, but also some of the cultures, human interests, and public opinions that exist only in the perceptions of people. Accordingly, extracting semantic relatedness from news articles can help us understand the diverse relations among cities. The proposed framework is based on the assumption that co-occurrence suggests relatedness, and aims to identify the main topics of the news articles that contain city co-occurrences. A core component of this framework is a supervised natural language processing model, Labeled Latent Dirichlet Allocation, which can automatically "read" a large number of news articles and compute likelihood scores for different topics. We described the overall structure of the framework as well as its individual modules. This framework was applied to a news article dataset with more than 500,000 news articles containing the top 100 cities in the contiguous U.S. to examine their semantic relatedness from 2006 to 2015. Based on the experiment result, we provide exploratory visualizations on the multiple city networks under different topics and the temporal variations of the semantic relatedness over the years. The proposed framework can be used in city network research to perform large-scale content analysis for understanding city relations (Pred 1980, Taylor 1997, Beaverstock et al. 2000, Taylor and Derudder 2016). The framework can also be generalized to other datasets and other topics depending on the application contexts. For example, in geographic information retrieval, one can train the framework using the news articles whose topics are similar to those of the actual data repository to be searched. In addition, the computational framework can be applied to not only news articles but also other types of textual documents, such as historical archives. Therefore, we can also use it to explore the evolutions of the semantic relatedness between some ancient cities.

A second contribution of this paper is the distance decay analysis of the extracted semantic relatedness. This analysis extends a previous study from Liu et al. (2014) who discovered a weak distance decay effect (with 0.2 as the friction factor) of the general place relatedness using the news articles from an outlet different from ours. Our hypothesis is that the semantic relatedness under different topics may have sometimes largely varied distance decay effects. To verify this hypothesis, we analyzed the semantic relatedness (under 17 topics) between cities along with their geodesic distances, and found that the general relatedness (i.e., without differentiating the topics) has a friction factor 0.23, which is very close to the value of 0.2 discovered by Liu et al. (2014). However, the semantic relatedness show varied friction factors, ranging from 0.03 ("Human Interest") to 0.61 ("Disaster, Accident and Emergency Incident"). This finding suggests that geographic distance has non-uniform impacts on place relatedness under different semantic topics.

This research can be extended in several directions. First, news article data from other outlets can be included to further examine the varied distance decay effects of semantic relatedness. Currently, our experiment used only the news articles from The Guardian, whose open API allows us to retrieve the full texts of all news articles of city pairs. This potential data bias does not affect the effectiveness of the proposed computational framework, but might introduce bias to the result of our distance decay analysis. Thus, further analysis can be performed when more news article data become openly available. This data limitation, however, should not be confused with the easy accessibility of news articles, since one can usually purchase digital news article data from agencies (Guardian data are open and free). Second, we can extend the current LLDA model with additional natural langue processing methods to acquire deeper understanding of the topics of news articles. So far, the proposed framework extracts the main topics of an entire news article. However, within one article, separate paragraphs may discuss related but probably also different topics, and therefore we can consider assigning topics to cities based on individual paragraphs. Further, the word distance between the co-occurring cities can be included to quantify the weight of relatedness, as suggested by Geiß et al. (2015). These additional methods can help us gain finer information on the semantic relatedness between cities, and thus are worth further research. While this work has limitations, we hope that the proposed computational framework and the distance decay analysis can make modest contributions to extracting and understanding semantic relatedness between cities.

**Acknowledgement**

The authors would like to thank Dr. David O'Sullivan and the four anonymous reviewers for their constructive suggestions and comments. This work was partially supported by the U.S. National Science Foundation (Award No. 1416509).